\documentclass[5p,times]{elsarticle}
\pdfoutput=1
\usepackage{subfig} 
\usepackage{graphicx}
\usepackage{float}
\usepackage{threeparttable}    
\usepackage{tabu}                     
\usepackage{multirow}                 
\usepackage{multicol}                 
\usepackage{multirow}                
\usepackage{float}                    
\usepackage{makecell}                 
\usepackage{booktabs}     
\usepackage{amssymb}
\usepackage{amsmath}
\usepackage[T1]{fontenc}

\usepackage{amsmath}            
\usepackage{epstopdf}           
\usepackage{flushend}           

\usepackage[figuresright]{rotating}
\usepackage{makecell}

\begin{document}
\begin{sloppypar}
\begin{frontmatter}

\title{LCAUnet: A skin lesion segmentation network with enhanced edge and body fusion}

\author[First]{Qisen Ma}
\author[First]{Keming Mao}

\author[First]{Gao Wang}


\author[Second,Third]{Lisheng Xu}

\author[Fourth]{Yuhai Zhao}

\cortext[cor1]{corresponding: Keming Mao}

\address[First]{Software College, Northeastern University, Shenyang, China}

\address[Second]{College of Medicine and Biological and Information Engineering, Northeastern
University, Shenyang, China}

\address[Third]{Engineering Research Center of Medical Imaging and Intelligent Analysis, Ministry of Education, Shenyang, China}

\address[Fourth]{College of Computer Science and Engineering, Northeastern
University, Shenyang, China}

\begin{abstract}

Accurate segmentation of skin lesions in dermatoscopic images is crucial for the early diagnosis of skin cancer and improving the survival rate of patients. However, it is still a challenging task due to the irregularity of lesion areas, the fuzziness of boundaries, and other complex interference factors. In this paper, a novel LCAUnet is proposed to improve the ability of complementary representation with fusion of edge and body features, which are often paid little attentions in traditional methods. First, two separate branches are set for edge and body segmentation with CNNs and Transformer based architecture respectively. Then, LCAF module is utilized to fuse feature maps of edge and body of the same level by local cross-attention operation in encoder stage. Furthermore, PGMF module is embedded for feature integration with prior guided multi-scale adaption. Comprehensive experiments on public available dataset ISIC 2017, ISIC 2018, and PH2 demonstrate that LCAUnet outperforms most state-of-the-art methods. The ablation studies also verify the effectiveness of the proposed fusion techniques.

\end{abstract}


\end{frontmatter}

\section{Introduction}

 Skin cancer is one of the most lethal cancer types. Based on the data from the American Cancer Society, there are about more than 99,000 new cases of melanoma by the end of 2022, with an associated mortality rate of 7.66$\%$ (7,650 cases) \cite{siegel2022cancer}. Early diagnosis and treatment of skin cancer can improve survival rates by up to 90$\%$ \cite{ge2017skin}. Skin lesion segmentation from dermatoscopic images plays a key role for this problem \cite{vestergaard2008dermoscopy}. In current clinical practice, dermatologists perform this task manually \cite{haenssle2018man}, which is a tedious, time-consuming, experience-dependent. Consequently, there is growing interest in exploring automated methods for skin lesion segmentation.
 
\begin{figure*}[htbp] 
		\centering
            {\includegraphics[scale=0.7]{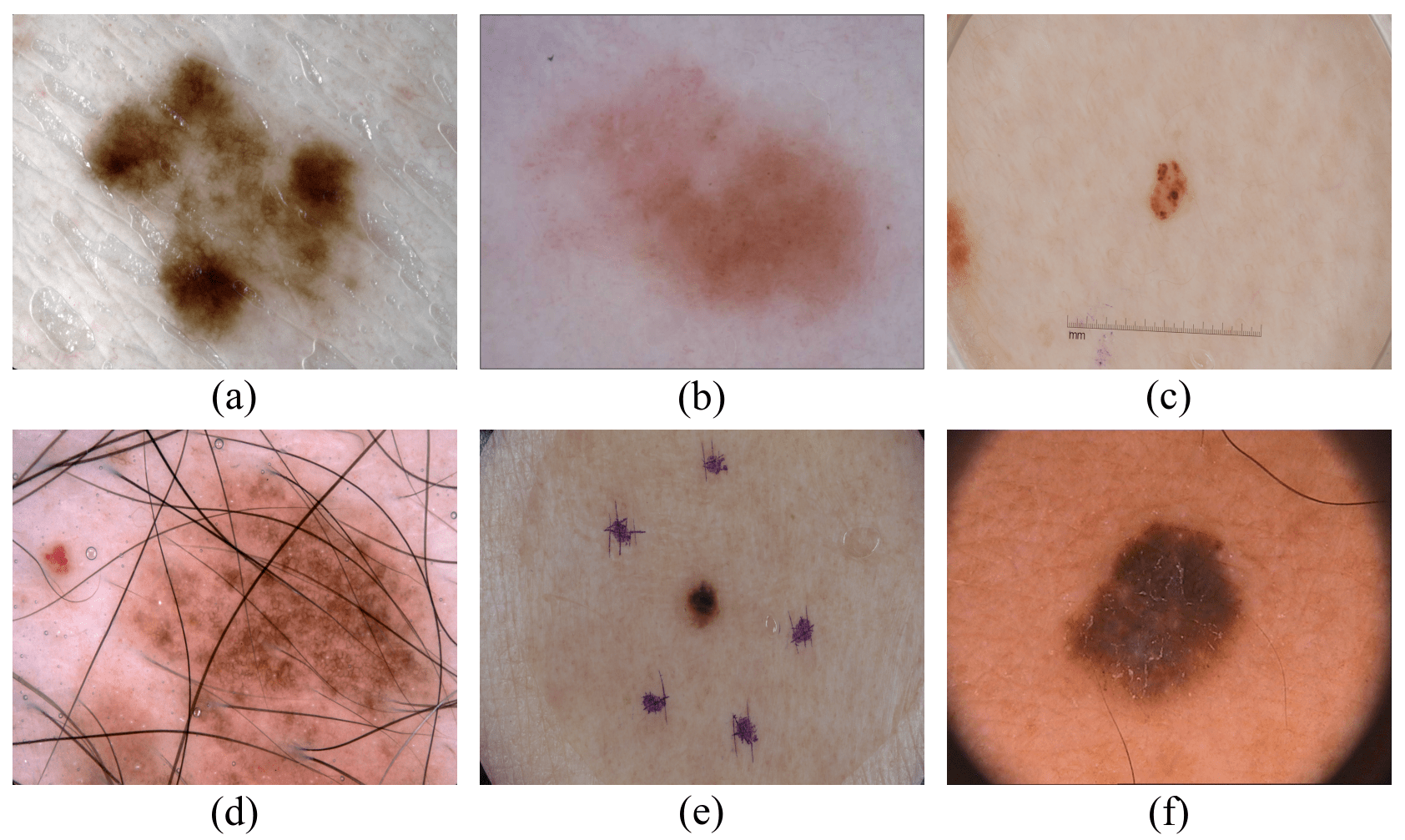}}	
		\caption{Examples of challenging in skin lesion segmentation: (a) irregular shape; (b) blurry boundary with low contrast; (c) small size; (d) hairs’ interference; (e) diagnostic marks’ interference; (f) circular field of view. }
        \label{sample}
	\end{figure*}

However, automatically differentiating lesions from healthy skin is a challenging task. On one hand, skin lesions are irregular in proportion, shape, location, and size, and are surrounded by blurry boundaries. On the other hand, dermoscopic images are often interfered by surrounding hair, ruler markings, and stained areas, which further complicates skin lesion segmentation, as shown in Fig. \ref{sample}. Traditional researches mainly focused on classical machine learning methods. Histogram thresholding methods segment skin damage areas from surrounding tissues by setting one or more thresholds \cite{silveira2009comparison, emre2013lesion, peruch2013simpler}. Clustering methods classify skin lesion areas by manually selecting features such as color, damage shape, and texture manually \cite{silveira2009comparison, garnavi2010automatic}. However, these methods relied heavily on the quality of the manually selected features, which are often insufficient to represent lesion information, leading to inferior performance.

Due to the development of deep learning technology in medical image segmentation, researchers have started to apply deep learning methods based on convolutional neural networks (CNNs) to skin lesions automatic segmentation. By directly extract prominent features from images without manual intervention, more richness representations can be learned and better model performance is obtained. A series of CNNs methods with U-shaped architecture have been paid more attentions, including Unet \cite{ronneberger2015u}, UNet++ \cite{zhou2018unet++}, ResUNet \cite{zhang2018road}, AttU-Net \cite{oktay2018attention}, etc. While CNNs have demonstrated effectiveness in skin lesion segmentation, they are limited in capturing long-range dependencies due to the inherent locality of convolutional operators, which may limit their capability to further segmentation.

Recently, the Transformer architecture, which employs self-attention mechanisms to capture global dependencies, has gained remarkable results on various visual tasks \cite{dosovitskiy2020image,liu2021swin}. Some studies explore its ability in medical image processing. Cao et al. \cite{cao2023swin} replaced the encoder and decoder in U-Net with Swin Transformer blocks and proposed Swin-UNet. He et al. \cite{he2022fully} integrated the Transformer with spatial pyramids to construct the hierarchical Fully Transformer Network model. Despite the Transformer's capacity to capture long-range dependencies via self-attention mechanisms, it cannot capture image locality and translational invariance effectively, leading to difficulty in accurately segmenting skin lesion boundaries.

Meanwhile, with the exceptional performance of fusing different types of features, researchers have also explored the potential of multi-branch architectures in medical image processing. Some studies utilized multi-branch networks to enhance the multi-scale feature representation. Valanarasu et al. \cite{valanarasu2021kiu} employed an undercomplete and an overcomplete encoder to address the problem of inaccurate segmentation of small structures caused by single encoder structure. Similarly, Lin et al. \cite{lin2022ds} utilized a dual-branch Swin Transformer encoder to extract multi-scale feature representations by feeding patches of different sizes. In addition, some studies used different branches to process different modalities of data and then fused the results. For instance, Zhu et al. \cite{zhu2023brain} employed a dual-branch approach to separately process magnetic resonance images of brain tumors with different modality, which outperformed single branch model. These studies have demonstrated the capacity of multi-branch models in tackling complex problems.

However, most existing deep learning-based methods for skin lesion segmentation mainly focus on lesion body while ignore the importance of edge information. For the few studies that considered both edge and body we reviewed,  Kuang et al. \cite{kuang2021bea} separated edge and body features through an element-wise subtraction operation, which resulted in rough edge feature detection. Similarly, Yang and Yang \cite{yang2023cswin} adopted the detected edge feature map as an additional constraint. These previous works did not fully leveraged the important role of edge features in skin lesion segmentation.

To address these issues, we propose a novel U-shaped network called LCAUnet based on the enhanced fusion of edge and body information. Specifically, LCAUnet consists of three main modules: a dual-branch encoder, a local cross-attention feature fusion module LCAF, a prior guided multi-scale fusion module PGMF. The dual-branch encoder is constructed to simultaneously extracts edge and body information. In detail, a lightweight CNNs branch integrated pixel-wise convolution \cite{su2021pixel} and a hierarchical Transformer branch based on SwinTransformer\cite{liu2021swin} are used for extracting edge and body features, respectively. The LCAF is designed by fusing features that are close in position between two branches, and it could accurately fuses cross-modal features while reducing the computational complexity. The PGMF module adopts the prior knowledge in different scale for better feature fusion. The proposed LCAUnet network is evaluated on three publicly available skin lesion datasets: ISIC2017, ISIC2018, and PH2. Ablation experiments and comparisons with state-of-the-art methods demonstrate the effectiveness of the proposed LCAUnet module.

Our contributions can be summarized as follows:
\begin{itemize}
  \item [\qquad(1)] 
A novel skin lesion segmentation network LCAUnet is proposed based on enhanced edge and body fusion. It is more capable of integrating complementary features and  convenient to handle irregular and difficult-to-detect boundary in skin lesions. 

  \item [\qquad(2)]
LCAF module is designed for edge and body fusion by local cross-attention operation, which is more effective in feature fusion. 

  \item [\qquad(3)]
PGMF module is adopted with prior guided multi-scale fusion, which integrates features of different scales with high-level knowledge modulation.

  \item [\qquad(4)]
Extensive experiments on publicly available datasets are conducted to evaluate LCAUnet. The results show that the proposed model achieve superior performance.

\end{itemize}

The rest of this paper is organized as follows. In Section 2, we describe the related works. The LCAUnet model is given in Section 3 in detail. Section 4 presents the experimental evaluation. Finally, Section 5 concludes this paper.

\section{Related works}

\subsection{CNNs-Based methods}

CNNs is a powerful network structure which extracts feature directly from raw images through multiple convolutional layers. The Fully Convolutional Network (FCN) \cite{long2015fully} is a classic work in the field of semantic image segmentation. Yuan et al. \cite{yuan2017automatic} proposed a 19-layer deep convolutional neural network (DCNN), which achieved fully automated segmentation of skin lesions through convolution and deconvolution operations. Inspired by FCN, Ronneberger et al. \cite{ronneberger2015u} proposed a U-shaped architecture network (U-Net) for biomedical image segmentation, which achieved excellent performance. Subsequently, a series of methods based on U-Net have been designed. Attention UNet \cite{oktay2018attention} suppressed the input of irrelevant regions by adding soft attention gates before the skip-connection, while highlighting salient features that are useful for specific tasks. Taghanaki et al. \cite{taghanaki2019select} replaced the skip-connections with a select-attend-transfer(SAT) gate, and it improved the model accuracy while reduced its memory usage by channel selection. Wu et al. \cite{wu2020automated} used an Adaptive Dual Attention Module before the skip-connection to parallel perform two global context modeling operations, which are often ignored. Furthermore, DSM \cite{zhang2019dsm} added side-output layers to the decoder part of the network to aggregate features from all levels. Multi-stage models were also 
studied. Bi et al. \cite{bi2017dermoscopic} proposed the parallel integration method in the multi-stage fully convolutional network (mFCN). Tang et al. \cite{tang2019multi} constructed a multi-stage U-Net (MS-UNet) based on deep supervision learning strategies to further improve model performance. Jha et al. \cite{jha2020doubleu} further polished the model's accuracy by stacking two U-Net structures in sequence.

In addition, several methods used generative adversarial networks (GANs) to explore latent representations \cite{bi2019improving, lei2020skin, sarker2021slsnet}. However, although the above methods have contributed much to the progress of skin lesions segmentation, technical bottlenecks in the segmentation of skin diseases still exist due to the inability to extract the global context information.



\subsection{Transformer-Based methods}

Recently, Transformer, a self-attention model derived from natural language processing, has attracted widespread attentions in image classification, semantic segmentation and object detection. Dosovitskiy et al. \cite{dosovitskiy2020image} proposed  Vision Transformer model (ViT). It divided images into non-overlapping 16x16 patches and achieved performance comparable to other state-of-the-art methods using convolution techniques. Liu et al. \cite{liu2021swin} devised a hierarchical Swin Transformer and achieved cross-window information exchange with higher performance while reducing computational complexity with sliding window strategy.

For its outstanding ability, researchers begun to explore Transformer in semantic segmentation. Some works attempted to use a vanilla transformer for semantic segmentation. E. Xie et al. \cite{xie2021segformer} proposed a simple but effective semantic segmentation model, SegFormer, which adopted lightweight multi-layer perceptron decoders. Cao et al. \cite{cao2023swin} replaced the encoder and decoder in U-Net with Swin Transformer blocks to establish Swin-UNet. He et al. \cite{he2022fully} combined the transformer with the spatial pyramid to construct a layered Fully Transformer Network model. Some works embedded transformer blocks in traditional CNNs architectures to enhance semantic segmentation. Transunet \cite{chen2021transunet} globally modeled low-level CNNs features by embedding transformer blocks between the encoder and decoder. MCTrans \cite{ji2021multi} enhanced features and extracted semantic information by passing multi-layer scale features extracted by CNNs into transformer blocks. However, although transformer-based models could effectively extract global semantic information, they have limitations in handling fine-grained segmentation.

\subsection{Multi-Branch methods}


Researchers tried to use multi-branch networks to enhance feature representation and fusion of different types and proved the superiority of multi-branch networks in handling complex problems. Some studies employed multi-branch networks to enhance the extraction of multi-scale features. Valanarasu et al. \cite{valanarasu2021kiu} used both an undercomplete and an overcomplete encoder to address single encoder structures’ inaccuracy in segmenting small structures. Lin et al. \cite{lin2022ds} fed different-sized image patches into a dual-branch Swin Transformer encoder in order to extract multi-scale feature representations. 

Some researchers have also attempted to combine CNNs and transformers in order to capture both local context and long-range dependency. Wu et al. \cite{wu2022fat} and Zhang et al. \cite{zhang2021transfuse} constructed dual-stream encoders, using CNNs encoders for spatial correlation modeling and transformer encoders for global information capturing. The features collected by the two encoders were then fused through a fusion module to obtain the final segmentation result. 

Other studies used different branches to process different modalities of data and then fused the features. TransFusion \cite{bai2022transfusion} and DeepFusion \cite{li2022deepfusion} first employed different branches to extract features of point clouds and images, and then fused the features with a cross-attention module. In the field of medical image segmentation, Zhu et al. \cite{zhu2023brain} used two branches for magnetic resonance images of different brain tumor feature representation, experiment results demonstrated its effectiveness over traditional single-branch network method.

\begin{figure*}[htbp]
		\centering
            {\includegraphics[width=0.95\textwidth]{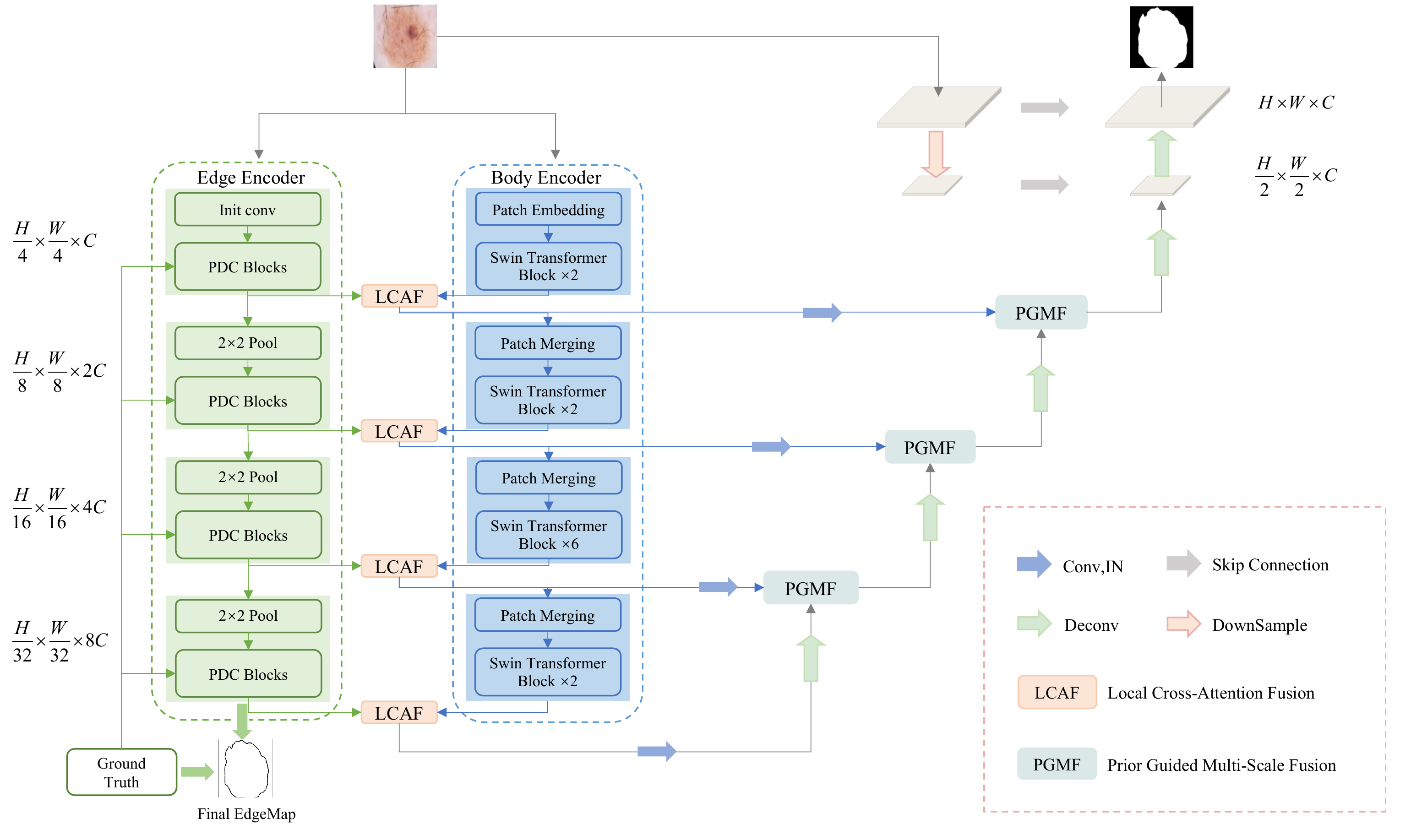}}	
		\caption{The proposed LCAUnet framework comprises a dual-encoder that incorporates both edge and body encoders to simultaneously extract edge and body features. Additionally, the framework includes an efficient fusion module, LCAF, which facilitates the fusion of edge and body features, and an AMSFF module that is embedded in the decoder to fully integrate adjacent scale features.}
        \label{LCAUnet}
	\end{figure*}

\section{Methodology}

The framework of LCAUnet is illustrated in Fig. \ref{LCAUnet}. The model mainly includes a dual-branch encoder for extracting edge and body features simultaneously, a cross-attention feature fusion module LCAF and a decoder which contains a PGMF module that fuse features with prior guided multi-scale knowledge. These components will be described in detail in the following sections.

\subsection{Dual-branch encoder}

The dual-branch encoder is composed of an edge encoder and a body encoder. The edge encoder adopts CNNs architecture, which uses PDC blocks based on pixel-wise difference convolution \cite{su2021pixel} and multi-scale feature maps to enhance the extraction of edge information. Meanwhile, the body encoder employs Transformer architecture, which captures body with rich semantic information by utilizing global attention.

\subsubsection{Edge encoder}

The edge features play fundamental and significant roles in image segmentation. Considering the problem of insufficient edge information extraction by conventional skin segmentation networks, a dedicated edge detection branch is employed in this study. As illustrated in Fig. \ref{LCAUnet}, the edge encoder consists of four stages which extract edge features at different levels. Each stage contains four PDC blocks for feature detection. To obtain hierarchical features, max-pooling layers are used to downsample the feature map between different stages. Notably, the init conv. in the first stage expands the original 3-channel image to C channels and reduces the feature maps to the size of 1/4 to ensure that the output size is equivalent to that of the body encoder.

The PDC block, comprises a depth-wise convolution layer, a ReLU layer, and a convolution layer with a kernel size of 1. Additionally, a residual connection is added to facilitate model training. Due to the lack of explicitly encoding gradient information, traditional convolutional networks have difficulty in focusing on extracting edge-related information. To overcome this problem, we introduce pixel-difference convolution \cite{su2021pixel}, to build a depth-wise convolution layer. Gradient information can be directly integrated into convolutional operations by pixel-difference convolution, and the edge feature is enhanced.

\begin{figure}[htbp]
		\centering
            {\includegraphics[width=0.5\textwidth]{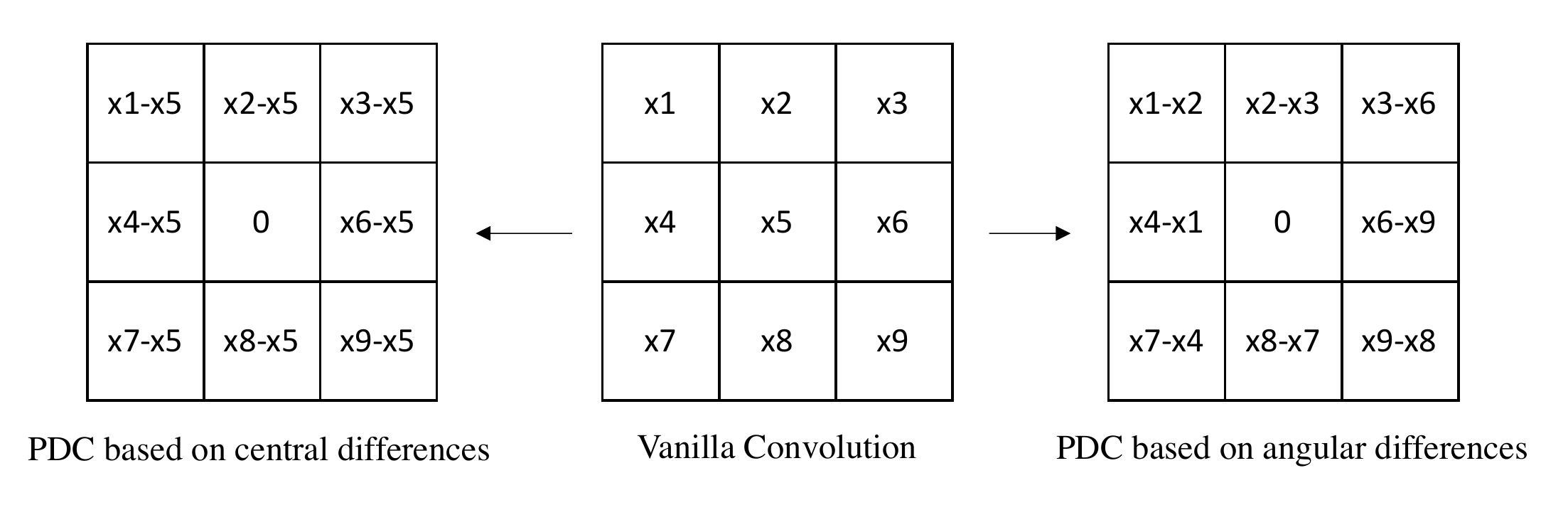}}	
		\caption{Description of generating PDC from
  vanilla convolution.}
        \label{PDC}
	\end{figure}
 
In contrast to vanilla convolution, the pixel difference convolution calculates the difference between pixel values of local features covered by the convolution kernel, rather than the original pixels (as illustrated in Fig. \ref{PDC}). The expressions for vanilla convolution and pixel difference convolution are provided below:

\begin{equation}
y=f(x,\theta )=\sum_{i=1}^{k\times k}w_i\cdot  x_i,\quad (vanilla\ convolution)
\end{equation}

\begin{equation}
y=f(\nabla x,\theta )=\sum_{(x_i,{x}'_i )\in P}^{}w_i\cdot  (x_i-x_i'),\quad (PDC)
\end{equation}

\noindent  where $w_i$ denotes the weights in the k × k convolution kernel, and $x_i$ and ${x}'_i$ denote the pixels covered by the kernel. The set $P = \{ (x_1, {x}'_1),  (x_2, {x}'_2),...,(x_m, {x}'_m)\}$ represents the collection of selected pixel pairs in the local area covered by the convolution kernel, and $m \le  k \times k$.

To further increase the constraint, we adopt the supervision strategy \cite{xie2015holistically}, by generating an edge map for the output feature of each stage and calculating the loss between the generated edge maps and the ground truth.

\subsubsection{Body encoder}

Transformer architecture is adopted for body encoder, since it is more effective to encode high-level feature representation with global long range modeling. Among various transformer models, Swin Transformer \cite{liu2021swin} is proficient in construct hierarchical features through a sliding window mechanism, and it is well-suited for segmentation of skin lesions with irregular shape. 

\begin{figure}[htbp]
		\centering
            {\includegraphics[width=0.3\textwidth]{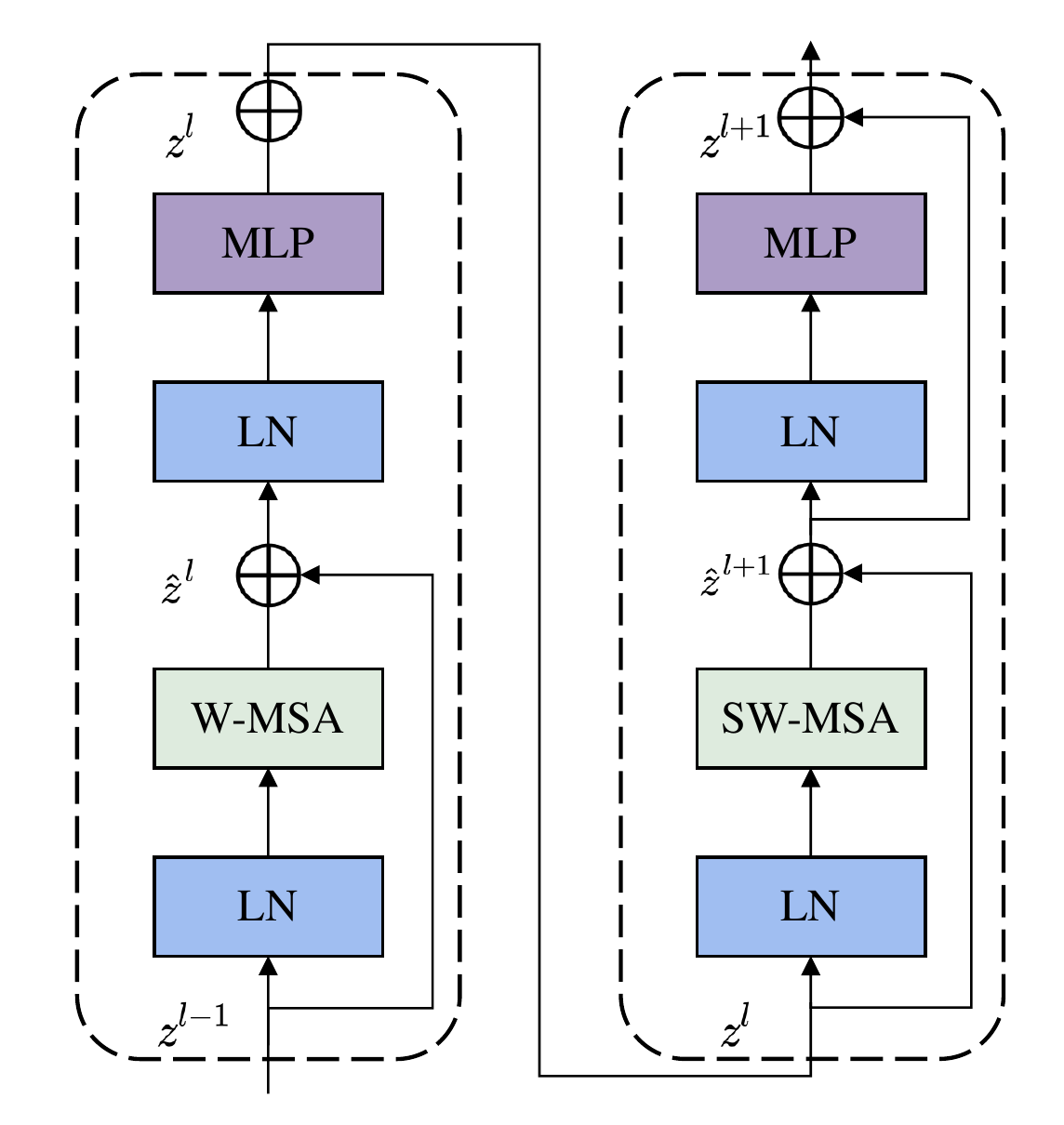}}	
		\caption{Two adjacent Swin Transformer blocks.}
        \label{SwinTransformer}
	\end{figure}

As shown in Fig. \ref{LCAUnet}, the Swin Transformer consists of four stages. The first stage is comprised of a patch embedding layer and two Swin Transformer blocks, which are utilized to perform feature encoding on the original image. Specifically, the input image is initially partitioned into M patches of size $P\times P$, and these patches are then reshaped into 1D vectors. Subsequently, these patches are then flattened and mapped to a C-dimensional space using a trainable linear projection. Learnable positional parameters are sequentially added to encode positional information of each patch. Finally, the sequence is fed into the Swin Transformer blocks. The latter three stages utilize patch merging and Swin Transformer blocks to progressively downsample the feature map while extracting higher-level features.

The Swin Transformer blocks in each stage comprises alternating arrangements of two different Swin Transformer blocks, as illustrated in Fig. \ref{SwinTransformer}. The first block is composed of Layer Normalization (LN), Window-based Multi-head Self-Attention (W-MSA), Multi-Layer Perceptron (MLP), and residual connections. The second type of block has a nearly identical structure, with the exception of using Shifted Window-based Multi-head Self-Attention (SW-MSA) in place of W-MSA.The above process can be represented as follows: 
\begin{flalign}
&\  \hat{z}^l=W\mbox{-}MSA(LN(z^{l-1}))+z^{l-1}  &\\
&\  z^{l}=MLP(LN(\hat{z}^{l}))+\hat{z}^{l} &\\
&\  \hat{z}^{l+1}=SW\mbox{-}MSA(LN(z^{l}))+z^{l}  &\\
&\   z^{l+1}=MLP(LN(\hat{z}^{l+1}))+\hat{z}^{l+1}  &
\end{flalign}
where ${z}^l$ denotes the input features at the $l$th
Swin Transformer block, $\hat{z}^l$ denotes the output features from  $l$th W-MSA or SW-MSA module.

Between each stage, Swin Transformer utilizes patch merging to perform down-sampling and gather contextual features. Patch merging operation merges adjacent $2\times 2$ patches into a larger patch to reduce the number of patches, and concatenates the dimensions of these patches to minimize information loss. By employing patch merging, the features can be downsampled $2\times$ each time. Assuming the size of  input image is $H \times W\times 3$, 
the output feature size of each stage is $\frac{H}{4}\times  \frac{W}{4}\times C$, $\frac{H}{8}\times  \frac{W}{8}\times 2C$,$\frac{H}{16}\times  \frac{W}{16}\times 4C$, $\frac{H}{32}\times  \frac{W}{32}\times 8C$, respectively.

Subsequently, the above extracted edge and body features are fed into the LCAF module for further fusion.

\subsection{LCAF module}

Cross-attention is a commonly used technique in the field of computer vision for fusing features from different modalities. Recent studies, such as TransFusion \cite{bai2022transfusion} and DeepFusion \cite{li2022deepfusion}, have shown impressive performance by merging features from point clouds and images. However, the conventional cross-attention method requires storing global information for each patch, resulting in high computational complexity. Moreover, for two distinct image modalities, the conventional approach is unable to exploit the one-to-one correspondence of pixels between modalities, making it difficult to achieve accurate matching and comprehensive fusion.

To tackle this problem, we propose an LCAF module that performs selective cross-attention operations on features that are in close proximity between the two image modalities. In this way, the edge and body features can be fused more accurately while reducing computation cost.

\begin{figure*}[htbp]
		\centering
            {\includegraphics[width=0.85\textwidth]{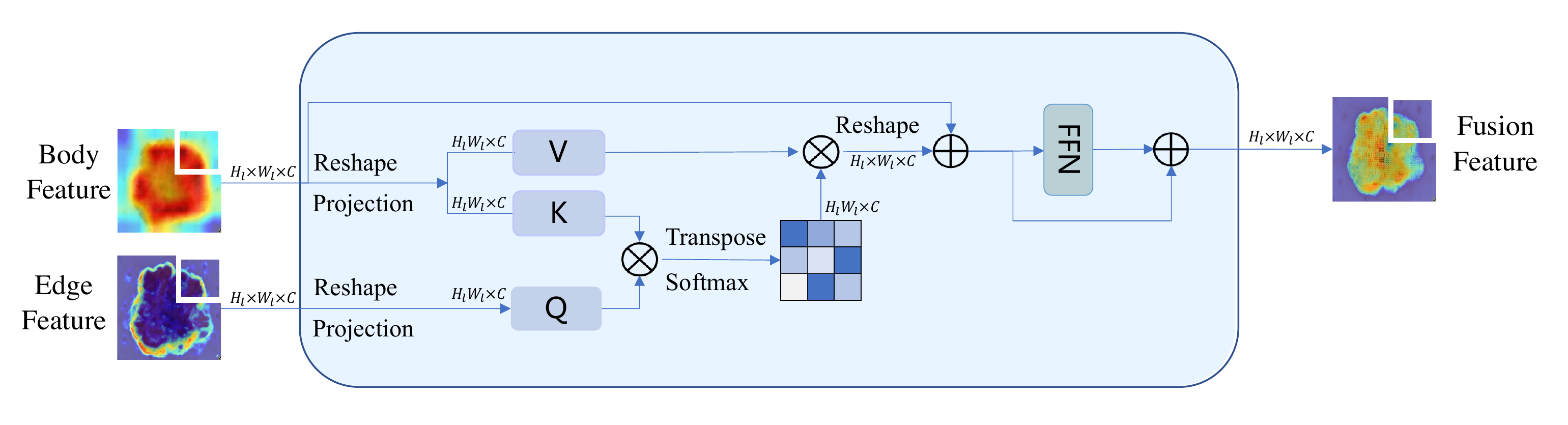}}	
		\caption{Structure diagram of LCAF.}
        \label{LCAF}
	\end{figure*}

Basically, LCAF module is a transformer block with local cross-attention, as shown in Fig. \ref{LCAF}. Assuming the input edge modality and body modality features have a size of $C\times H \times W$, then they are divided in to size of $C \times H_l \times W_l$(with window of $H_l \times W_l$) and reshaped to $H_l W_l \times C$. Linear projection is performed on the local features of the edge modality to obtain the query vector and on the local features of the body modality to obtain the key and value vectors. Next, attention score is calculated by taking the dot product of the query and key vectors. To ensure stability of the gradients, we divide the attention score by the square root of the dimensionality of the key vector. The resulting vector is then normalized through the softmax activation and multiplied by the normalized score to obtain the output vector. Finally,
the value vectors are weighted sum. The above local cross-attention(LCA) mechanism can be expressed as follows:

\begin{equation}
Query=X_{edge}W_Q,Key=X_{body}W_K,Value=X_{body}W_v
\end{equation}

\begin{equation}
LCA(X_f) = softmax(\frac{Query(Key)^T}{\sqrt{d_k} })Value
\end{equation}

\noindent where $X_{edge}$, $X_{body}$, $X_f$ denote edge modality features, body modality features, and the fused modality features, respectively. $W_Q$, $W_k$, $W_V$ are three learnable matrixs which denote the query matrix, the key matrix, and the value matrix, respectively.

Based on local cross attention, a multi-head operation is employed that projects identical query, key, and value vectors onto distinct $h$ subspaces within the original high-dimensional space while concurrently performing local cross-attention. Subsequently, multi-heads are concatenated for computing local cross-attention scores in diverse subspaces. During this process, the dimensionality of each vector is reduced to avoid overfitting. Moreover, representation from multiple subspaces can be better encoded. The specific formula for the M\mbox{-}LCA operation is provided below:
\begin{equation}
M\mbox{-}LCA(X_f) = X_f + Concat[LCA(X_f)_1, ... , LCA(X_f)_h]W_o
\end{equation}

\noindent where $W_O$ is a learnable matrix for output. The output of the LCAF module can be obtained through the following operation: 

\begin{equation}
X_f=FFN(M\mbox{-}LCA(X_f)) + M\mbox{-}LCA(X_f)
\end{equation}

\noindent where the feed-forward network(FFN) consists of two linear layers and utilizes the GeLu activation function to transform the feature space of M\mbox{-}LCA. Finally, residual connections are introduced to prevent network degradation.

Additionally, assuming that the features comprise $h\times w$ patches and each LCA-performing window contains $h_l\times w_l$ patches, the computational complexities of the conventional global cross-attention(GCA) and our proposed local cross-attention(LCA) are as follows:
\begin{align}
&\  \Omega (M\mbox{-}GCA) = 4hwC^2 + 2(hw)^2C  &\\
&\  \Omega (M\mbox{-}LCA) = 4hwC^2 + 2h_lh_w\cdot hwC &
\end{align}

\noindent where the computational complexity of the former is quadratic with patch number $hw$, while the latter is linear complexity with patch number $h_l\times w_l$. Compared with directly performing conventional global attention, LCAF module achieves effective fusion while reducing the computational complexity. 

The LCAF module is performed and the fused feature is incorporated into the next stage.

\subsection{Decoder with PGMF module}

To obtain pixel-level prediction results, a novel decoder structure is constructed with integration of multi-scale features.

Firstly, the fused features from each stage of the encoder are fed into a residual block for initial integration. The residual block consists of two $3\times 3$ convolution layers and is normalized by an instance normalization layer.

\begin{figure*}[htbp]
		\centering
            {\includegraphics[scale=0.4]{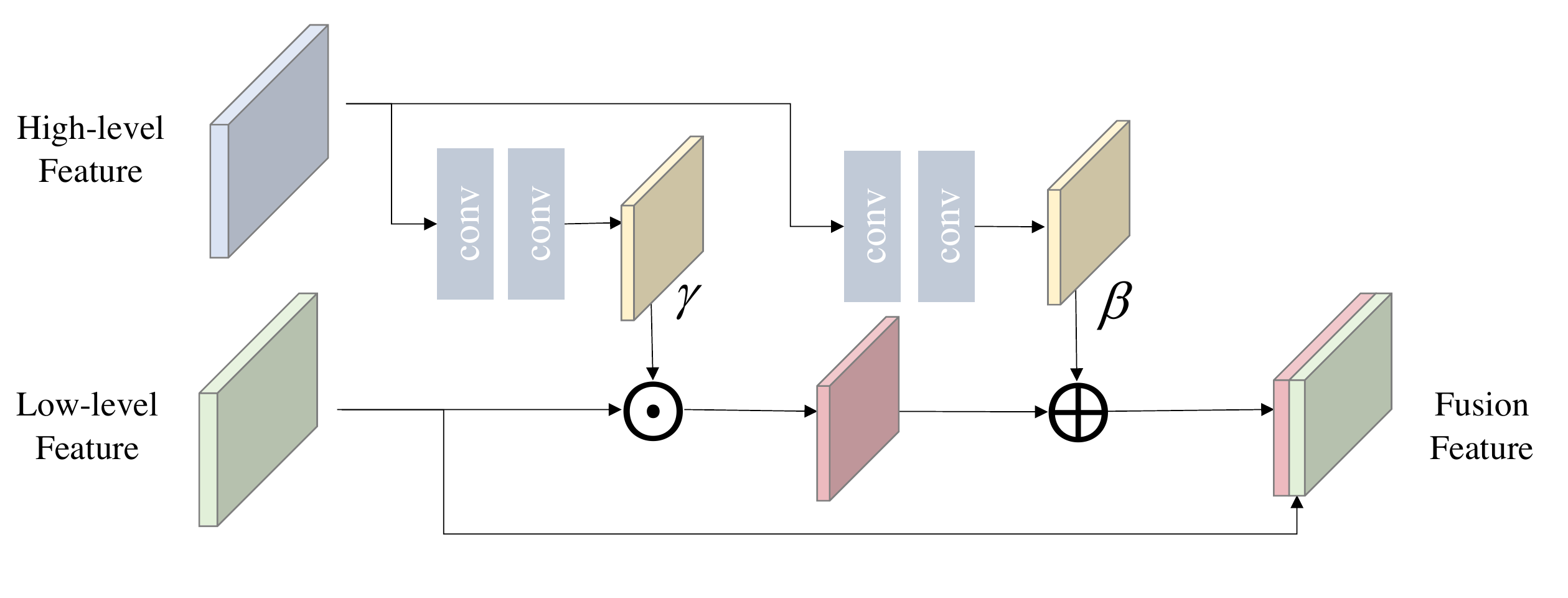}}	
		\caption{Structure diagram of PGMF.}
        \label{PGMF}
	\end{figure*}

Subsequently, the PGMF module is employed to fully integrate adjacent-scale features, which is superior to simple concatenation, as shown in Fig. \ref{PGMF}. The PGMF module is based on the Spatial Feature Transform \cite{wang2018recovering}, which uses high-level features as prior knowledge to modulate low-level features for advanced feature fusion. At the end of the PGMF module, the fused features are concatenated with low-level features to increase the redundancy of features, which helps to reduce the loss of fine-grained information during the upsampling.

Finally, through three PGMF modules, the output feature size is $\frac{H}{4} \times \frac{W}{4} $. It is detrimental for accurate pixel-level prediction since there is much loss of low-level feature by four-fold upsampling. Therefore, we adopt the approach proposed by Lin et al. \cite{lin2022ds}, which utilizes two convolutional blocks to downsample the input image and obtain low-level features with resolutions of $H \times W$ and $\frac{H}{2} \times \frac{W}{2} $. These two features are then concatenated with the outputs of PGMF module respectively.

\subsection{Loss function}

Since our encoder is designed to detect both edge and body features simultaneously, two loss components $L_{edge}$ and $L_{body}$, are employed.

\subsubsection{Edge supervision loss}

For edge detection, the majority of samples are negative, thus an annotator-robust loss \cite{liu2017richer} is employed.

We utilize this loss function on the edge maps generated by each stage of the encoder. For the i-th pixel in the j-th edge map, with a predicted result of $y^j_i$, its loss is computed as follows:
\begin{equation}
l^j_i= 
\begin{cases}
\alpha \cdot log(1-y^j_i) & y_i = 0 \\
0 & 0<y_i<\eta \\
\beta \cdot logy^j_i & other
\end{cases}
\end{equation}
where $y^j_i$ denotes the predicted value of the i-th pixel in the j-th edge map, and $\eta $ is a predefined threshold. If the pixel is annotated as positive by annotators with a proportion smaller than $\eta $ , it is considered to be a negative sample. $\beta $ represents the proportion of negative samples in the dataset. $\alpha =\lambda \cdot (1-\beta )$, where $\lambda$ is a hyperparameter used to balance positive and negative samples. The $L_{edge}$ is obtained by adding up the loss of each pixel, as follows: 
\begin{equation}
L_{edge}= {\textstyle \sum_{i,j}^{}} l^j_i
\end{equation}

\subsubsection{Body supervision loss}

The task of skin lesion segmentation suffers from a severe class imbalance problem. To address this, we adopt a combination of Binary Cross-Entropy Loss and Dice Loss \cite{milletari2016v} as our body loss. 

Binary Cross-Entropy Loss measures pixel-level prediction errors and is applicable to most semantic segmentation scenarios, which is given as below:
\begin{equation}
L_{Bce} = -  {\textstyle \sum_{i=0}^{N}}[(1-\hat{y}_i)ln(1-y_i) +\hat{y}_iln(y_i)] 
\end{equation}

Dice loss is a commonly used loss function in medical image segmentation, and it is also an effective way for imbalanced samples. The formula is given as follow:
\begin{equation}
L_{Dice} = 1 - 2\times \frac{2 {\textstyle \sum_{i=0}^{N}} y_i \hat{y}_i}{ {\textstyle \sum_{i=0}^{N}} (y_i + \hat{y}_i)}
\end{equation}

$L_{body}$ can be expressed in the following form:

\begin{equation}
L_{body}=\lambda _1L_{Bce} + \lambda _2L_{Dice}
\end{equation}

Combing $L_{edge}$ and $L_{body}$ with a weighted hyperparameter $\gamma$, the final loss $L$ can be computed as follows:

\begin{equation}
L=L_{body} + \gamma  \cdot L_{edge}
\end{equation}

\section{Experiments}

\subsection{Datasets}

In this paper, three publicly accessible datasets are used for performance evaluation.

$\bullet$ \textbf{ISIC2017 dataset}. The ISIC 2017 dataset \cite{codella2018skin} consists of 2000 training images, 150 validation images, and 600 test images, all of which have been manually annotated by professional dermatologists for segmentation tasks.Following Cheng et al. \cite{cheng2022resganet}, the colors of the images are first normalized by implementing the gray world algorithm.

$\bullet$ \textbf{ISIC2018 dataset}. The ISIC 2018 dataset \cite{codella2019skin} consists of 2594 RGB skin lesion images, including various types of skin lesions with different resolutions. Following the partitioning strategy used by Wu et al. \cite{wu2022fat}, we set the ratio of training set:validation set:test set as 7:1:2, resulting in 1815 images for the training set, 259 images for the validation set, and 520 images for the test set.

$\bullet$ \textbf{PH2 dataset}. The PH2 dataset \cite{mendoncca2013ph} comprises 200 RGB skin lesion images. We employ a partitioning strategy of 7:1:2 for the ratio of training set, validation set, and test set, respectively. Specifically, 140 images are randomly selected for the training set, 20 images for the validation set, and the remaining 40 images for the test set.

\subsection{Implementation details}

The proposed model is implemented with the PyTorch framework and all experiments are tested on a Nvidia Titan RTX 24G GPU. To improve the computational efficiency, we set the resolution of all training, validation, and test images to $224\times 224$. Additionally, to obtain better model initialization, we used PiDiNet \cite{su2021pixel} as the pre-trained model for the edge encoder and swin-tiny-patch4-window7-224 \cite{liu2021swin} for the body encoder.

The proposed model is trained using an AdamW optimizer with weight decay of 0.01. The initial learning rate is set to 0.01, and the ReduceLROnPlateau algorithm is employed for learning rate scheduling. A batch size of 24 and 80 training epochs are adopted. Weight parameters $\lambda _1$, $\lambda _2$  and $\gamma$ are set to 0.6, 0.4 and 0.2 respectively. 

To overcome the overfitting problem, follow by  Wu et al. \cite{wu2022fat},  various data augmentations are used, including randomly flipping horizontally and vertically, randomly rotating by angles between -15 and 15 degrees, and randomly changing the brightness and contrast within a certain range.


\subsection{Evaluation metrics}

Accuracy (ACC), Dice coefficient (Dice), Intersection over Union (IoU), Sensitivity (SE), and Specificity (SP) are adopted as evaluation metrics in this study.
\begin{flalign}
&\ \quad ACC = \frac{TP+FN}{TP+TN+FP+FN}  &\\
&\ \quad Dice = \frac{2\cdot TP}{2\cdot TP+FP+FN} &\\
&\ \quad IoU = \frac{TP}{TP+FP+FN} &\\
&\ \quad SE = \frac{TP}{TP+FN}  &\\
&\ \quad SP = \frac{TN}{TN+FP}  &
\end{flalign}
where TP (true positive) and TN (true negative) represent the numbers of correctly segmented skin lesion and background pixels, respectively, while FP (false positive) represents the number of background pixels wrongly labeled as skin lesion pixels, and FN (false negative) represents the number of skin lesion pixels wrongly predicted as background pixels. The values of all evaluation metrics range between 0 and 1, with a value closer to 1 indicating better segmentation results, and vice versa.

\subsection{Results on the ISIC 2017 dataset}

\begin{table}[htbp]\footnotesize
\renewcommand{\arraystretch}{1.5}
  \begin{center}
    \caption{Results of LCAUnet on ISIC2017 dataset compared with the latest methods.}
    \label{ISIC2017_table}
    \begin{tabular*}{\hsize}{@{}@{\extracolsep{\fill}}l l l l l l@{}} 
    \hline
      Method & Dice & SE &SP &ACC&IoU\\
      \hline
      U-Net \cite{ronneberger2015u} & 0.783 & 0.806 & 0.954 & 0.933 & 0.696\\
      UNet++ \cite{zhou2018unet++}& 0.832 & 0.830 & 0.965 & 0.925 &0.743\\
      Att U-Net \cite{oktay2018attention} & 0.808 & 0.800&0.978&0.915&0.717\\
      FocusNet \cite{kaul2019focusnet}& 0.832 & 0.767&\textbf{0.990}&0.921&0.756\\
      DoubleU-Net \cite{jha2020doubleu}& 0.845 & 0.841&0.967&0.933&0.760\\
      DAGAN \cite{lei2020skin}& 0.859 & 0.835&0.976&0.935&\textbf{0.771}\\
      TransUnet \cite{chen2021transunet}& 0.841 & 0.807&0.979&0.932&0.755\\
      FAT-Net \cite{wu2022fat}& 0.850 & 0.840&0.973&0.933&0.765\\
      ResGANet-MsASPP \cite{cheng2022resganet}& 0.862 & 0.842&0.950&0.936&0.764\\
      Ours(LCAUnet) & \textbf{0.866} & \textbf{0.852}&0.965&\textbf{0.940}&0.761\\
      \hline
    \end{tabular*}
  \end{center}

\end{table}

We compare the proposed LCAUnet with various state-of-the-art methods on the ISIC 2017 dataset. The results are presented in Table \ref{ISIC2017_table}. Among these methods, AttU-Net  achieves better performance than U-Net by using soft attention gates inserted before the skip connections to suppress irrelevant regions in the input image and highlight salient features that are useful for segmentation. DAGAN \cite{lei2020skin} further improves segmentation performance by integrating dense convolution and a double-discrimination (DD) module. Among these methods, LCAUnet achieves the highest scores on most metrics, with Dice, SE, SP, ACC, and IoU scores of 86.6$\%$, 85.2$\%$, 96.5$\%$, 94.0$\%$, and 76.1$\%$, respectively. Specifically, our proposed model ranks first on the two key metrics, ACC and Dice, respectively, which demonstrates that LCAUnet is highly competitive.

\begin{figure*}[htbp]
		\centering
            {\includegraphics[width=0.8\textwidth]{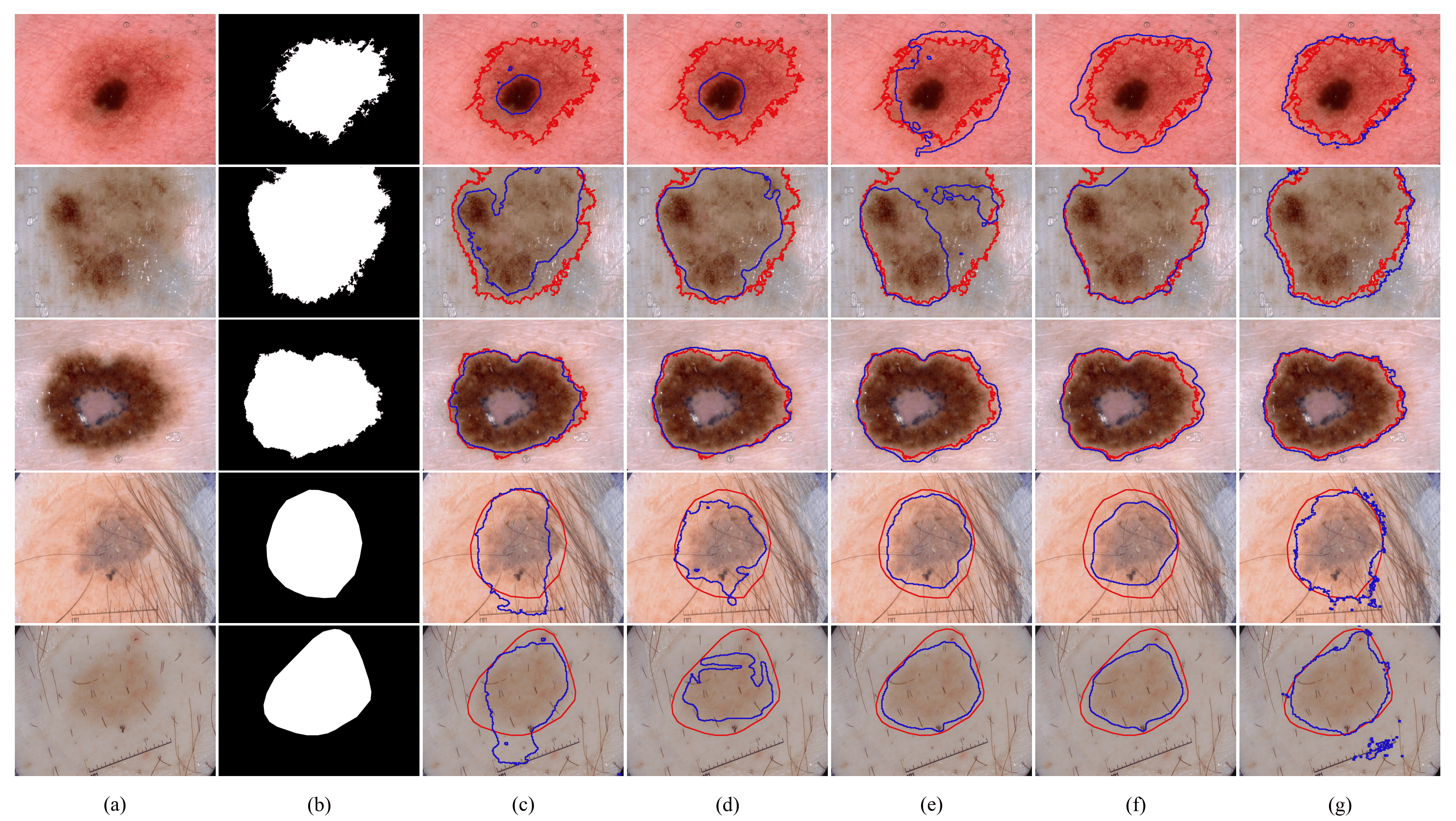}}	
		\caption{Visual comparison with different state-of-the-art methods on the ISIC 2017 dataset. (a) Input images. (b) Ground truth. (c) U-Net. (d) AttU-Net. (e) FAT-Net. (f) TransUnet. (g) Ours. The red outline and blue outline respectively indicate the ground truth and segmentation results.}
        \label{ISIC2017_visual}
	\end{figure*}

For further evaluation, we visualize and compare the segmentation results with several representative methods, including UNet, AttU-Net, FAT-Net, TransUNet, as shown in Fig. \ref{ISIC2017_visual}. LCAUnet outperforms the others and achieves the best segmentation results in diverse images. When the contrast between the foreground and background pixels in the image is low, U-Net and AttU-Net fails to accurately predict skin lesions. TransUnet and FAT-Net could achieve more accurate segmentation by combining the fine-grained features provided by CNNs and the global features provided by the transformer. However, these methods all show unsatisfactory performance in the edge region segmentation of skin lesions. Our proposed LCAUnet effectively extracts edge information through a powerful edge encoder, which greatly assists the segmentation, especially in cases where the boundaries of skin lesions are blurred.

\subsection{Results on the ISIC 2018 dataset}

\begin{table}[htbp]\footnotesize
\renewcommand{\arraystretch}{1.5}
  \begin{center}
    \caption{Results of LCAUnet on ISIC2018 dataset compared with the latest methods.}
    \label{ISIC2018_table}
    
    \begin{tabular*}{\hsize}{@{}@{\extracolsep{\fill}}l l l l l l@{}} 
    \hline
      Method & Dice & SE &SP &ACC&IoU\\
      \hline
      U-Net \cite{ronneberger2015u}& 0.855 & 0.880 & 0.970 & 0.940 & 0.773\\
      Att U-Net \cite{oktay2018attention}& 0.857 & 0.867 & 0.984 & 0.938 &0.776\\
      CPFNet \cite{feng2020cpfnet}& 0.877 & 0.895&0.966&0.950&0.799\\
      TransUnet \cite{chen2021transunet}& 0.850 & 0.858&\textbf{0.986}&0.945&0.809\\
      ERU \cite{nguyen2020skin}& 0.881 & 0.903 & 0.969 & 0.944 & 0.806 \\
      CKDNet \cite{jin2021cascade}& 0.878 & 0.906&0.970 &0.949&0.804\\
      FAT-Net \cite{wu2022fat}& 0.890 & 0.910&0.970 &0.958&0.820\\
      TMU-Net \cite{azad2022contextual}& 0.906 & 0.904&0.975&0.960&0.834\\
      Ours & \textbf{0.919} & \textbf{0.913} &0.978&\textbf{0.964}&\textbf{0.845}\\
      \hline
    \end{tabular*}
  \end{center}
\end{table}

We compare the proposed LCAUnet with 8 state-of-the-art methods on the ISIC 2018 dataset. The comparison results are shown in Table \ref{ISIC2018_table}. It can be observed that our method generally performs better than other comparative models in most metrics, and achieves the best results in two key metrics, ACC and Dice. Moreover, LCAUnet outperforms the latest model TMU-Net, with improvements of 1.31$\%$, 0.92$\%$, 0.34$\%$, 0.40$\%$, and 1.17$\%$ for metrics such as Dice, SE, SP, ACC, and IoU, respectively.

\begin{figure*}[htbp]
		\centering
            {\includegraphics[width= 0.8\textwidth]
            {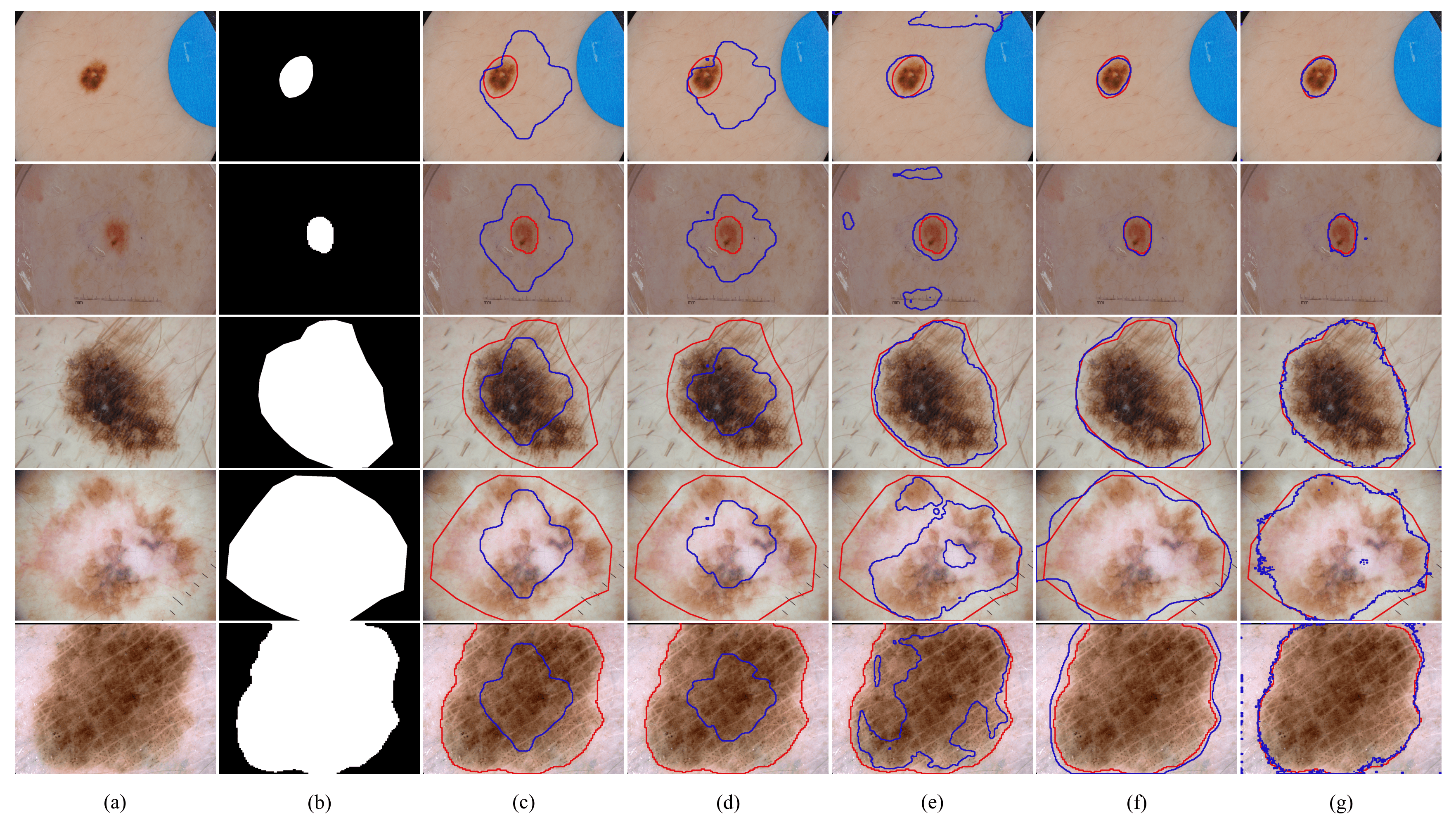}}		
		\caption{Visual comparison with different state-of-the-art methods on the ISIC 2018 dataset. (a) Input images. (b) Ground truth. (c) U-Net. (d) AttU-Net. (e) FAT-Net. (f) TransUnet. (g) Ours. The red outline and blue outline respectively indicate the ground truth and segmentation results.}
        \label{ISIC2018_visual}
	\end{figure*}

Similarly, we select several representative methods, including UNet, AttU-Net, FAT-Net, TransUNet, and LCAUnet, to visualize and compare the segmentation results of the ISIC 2018 dataset. As shown in Fig. \ref{ISIC2018_visual}, our method obtains better than other comparative models in terms of the visual effectiveness of skin lesion segmentation. Among these methods, UNet and AttU-Net are capable of identifying skin lesion areas, but their segmentation performance are poor for skin regions of different sizes. Despite the fusion of fine-grained information of CNN and the global information of the transformer through squeeze and excitation (SE) operation, FAT-Net has poor anti-interference ability and performs poorly in the segmentation of irregular objects. Although TransUNet performs better than previous models in segmentation results, it still has shortcomings in fine-grained edge segmentation. Compared with the other four competitors, our method achieves the best performance in skin lesion segmentation. Even in uncertain lighting conditions and with interference, our model could still accurately segment skin lesions of different scales and irregular shapes.

\subsection{Results on the PH2 dataset}

Finally, we conduct additional experiments on the pH2 dataset to demonstrate the model generalization ability. In contrast to the ISIC series dataset that comprises thousands of skin lesion images, the PH2 dataset provided by Pedro Hispano Hospital consists of only 200 images. Therefore, the comparison results on the PH2 dataset can evaluate the model's generalization ability on small datasets. We compared our LCAUnet with 8 state-of-the-art methods, and the comparisons are presented in Table \ref{PH2_table}.

\begin{table}[htbp]\footnotesize
\renewcommand{\arraystretch}{1.5}
  \begin{center}
    \caption{Results of LCAUnet on PH2 dataset compared with the latest methods.}
    \label{PH2_table}
    \begin{tabular*}{\hsize}{@{}@{\extracolsep{\fill}}l l l l l l@{}} 
    \hline
      Method & Dice & SE &SP &ACC&IoU\\
      \hline
      U-Net \cite{ronneberger2015u}& 0.894 & 0.913 & 0.959 & 0.923 & 0.841\\
      Att U-Net \cite{oktay2018attention}& 0.900 & 0.921 & 0.964 & 0.928 &0.858\\
      EDLM \cite{goyal2019skin}& 0.918 & 0.924&0.948&0.945&0.853\\
      DSNet \cite{hasan2020dsnet}& 0.920 & 0.960&0.961&0.948&0.872\\
      iFCN \cite{ozturk2020skin}& 0.932 & \textbf{0.961}&0.959&0.961&0.876\\
      MB-DCNN \cite{xie2020mutual}& 0.933 & 0.954&0.953&0.959&0.871\\
      FAT-Net \cite{wu2022fat}& 0.944 & 0.944&0.974&0.970&0.896\\
      APT-Net \cite{zhang2022apt}& 0.946 & 0.940&0.979&0.965&0.899\\
      Ours & \textbf{0.958} & 0.954&\textbf{0.982}&\textbf{0.973}&\textbf{0.918}\\
      \hline
    \end{tabular*}
  \end{center}
\end{table}

Among these methods, DSNet constructs a lightweight network by using depthwise separable convolutions instead of standard convolutions, achieving better performance than U-Net while reducing model parameters. By incorporating color processing and lesion center detection modules into a fully convolutional network, iFCN further improves the accuracy of skin lesion segmentation. MB-DCNN combines classification and segmentation networks, utilizing the former to precisely locate and diagnose skin lesions, and the latter to obtain more accurate lesion segmentation based on prior information from the classification network. APT-Net combines CNN and Transformer models to capture both local and global information, and uses adaptive positional encoding to generate token position information, further enhancing the model's accuracy. However, these methods do not sufficiently emphasize edge features, resulting in poor segmentation performance of lesion edges and hindering further improvement in accuracy. In contrast, our proposed LCAUnet, constructed with a dual-branch encoder, effectively captures and fuses both edge and body information to achieve superior segmentation performance. As shown in Table \ref{PH2_table}, LCAUnet model achieves the highest scores of 95.8$\%$, 98.2$\%$, 97.3$\%$, and 91.8$\%$ in Dice, SP, ACC, and IoU metrics, respectively, outperforming other methods.In addition, it is evident that these models' evaluation performance on the PH2 dataset exceeds that of the ISIC2017 and ISIC2018 datasets. It is hypothesized that this may be attributed to the relative simplicity of the PH2 dataset, with more regular skin lesion areas and less interference.

\begin{figure*}[htbp]
		\centering
            {\includegraphics[width=0.8\textwidth]
            {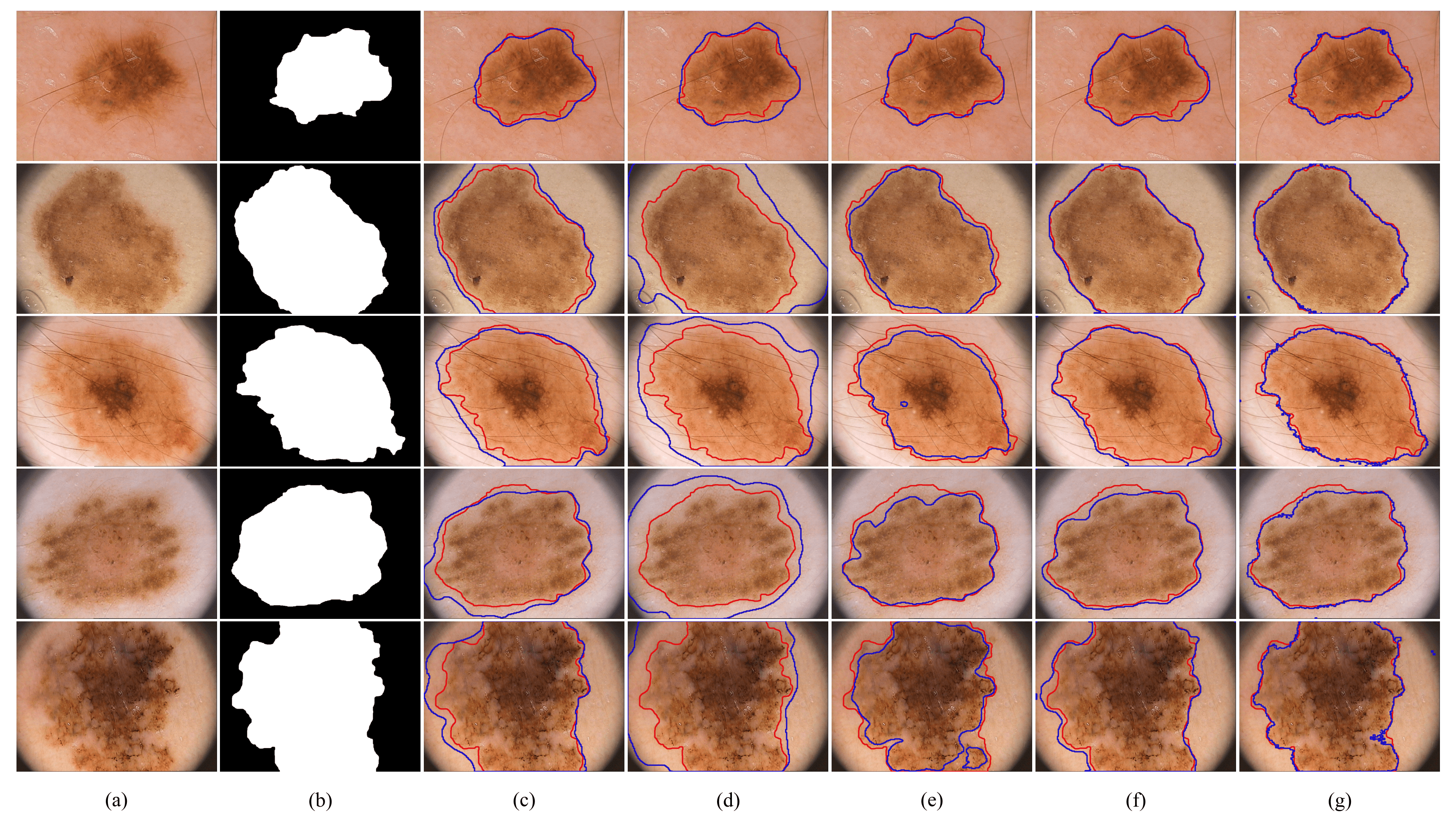}}		
		\caption{Visual comparison with different state-of-the-art methods on the PH2 dataset. (a) Input images. (b) Ground truth. (c) U-Net. (d) AttU-Net. (e) FAT-Net. (f) TransUnet. (g) Ours. The red outline and blue outline respectively indicate the ground truth and segmentation results.}
        \label{PH2_visual}
	\end{figure*}

\begin{table*} \footnotesize
\renewcommand{\arraystretch}{1.5}
    \centering
    \caption{The ablation experiment results of LCAUnet on the ISIC 2017, ISIC 2018 and PH2 datasets.}
    \label{ablation_table}
    \begin{tabular*}{\hsize}{@{}@{\extracolsep{\fill}}lllllllllll@{}}
        \Xhline{0.4pt}
         & \multicolumn{3}{l}{ISIC 2017} &\multicolumn{3}{l}{ISIC 2018}  &\multicolumn{3}{l}{PH2}\\
        \cmidrule(r){2-4}
        \cmidrule(r){5-7}
        \cmidrule(r){8-10}
        Method&ACC&IoU&Dice&ACC&IoU&Dice&ACC&IoU&Dice\\
        \Xhline{0.4pt}
        
        Baseline & 0.796 & 0.920 & 0.672 & 0.886 & 0.936 & 0.796 & 0.953 & 0.855 & 0.922\\
        Baseline + PGMF & 0.821 & 0.926 & 0.698 & 0.897 & 0.948 & 0.807 & 0.957 & 0.870 & 0.931\\
        Baseline + EE & 0.848 & 0.937 & 0.740 & 0.912 & 0.959 & 0.829 & 0.961 & 0.882 & 0.940\\
        Baseline + EE + LCAF & 0.858 & 0.938 & 0.746 & 0.914 & 0.962 & 0.837 & 0.968 & 0.904 & 0.952\\
        Baseline + EE + LCAF + PGMF(Ours) & \textbf{ 0.866} & \textbf{0.940} & \textbf{0.761} & \textbf{0.919} & \textbf{0.964} & \textbf{0.845} & \textbf{0.973} & \textbf{0.918} & \textbf{0.958}\\
 
        \Xhline{0.4pt}
    \end{tabular*} 
\end{table*}

\begin{figure*}[htbp]
  \centering
  \includegraphics[width=0.8\textwidth]{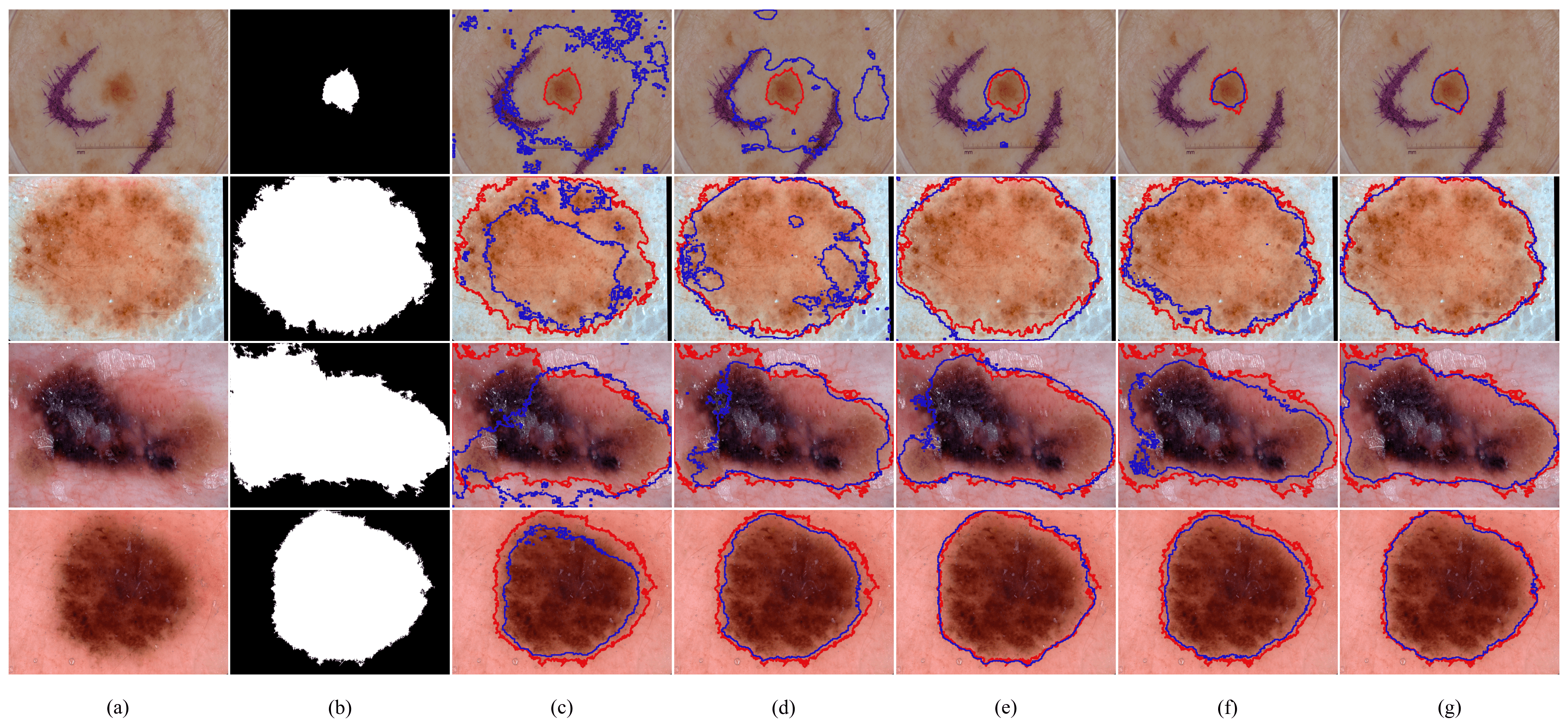}
  \caption{Visualization of ablative analysis for key modules in LCAUnet.(a) Input images. (b) GroundTruth. (c)Baseline. (d)Baseline+PGMF. (e) Baseline+EE(EdgeEncoder). (f) Baseline+EE+LCAF. (g) Baseline+EE+ LCAF+PGMF(Ours).The red outline and blue outline respectively indicate the ground truth and segmentation results.}
  \label{ablation_visual}
\end{figure*}

\begin{figure*}[htbp]
  \centering
  \includegraphics[scale=0.7]{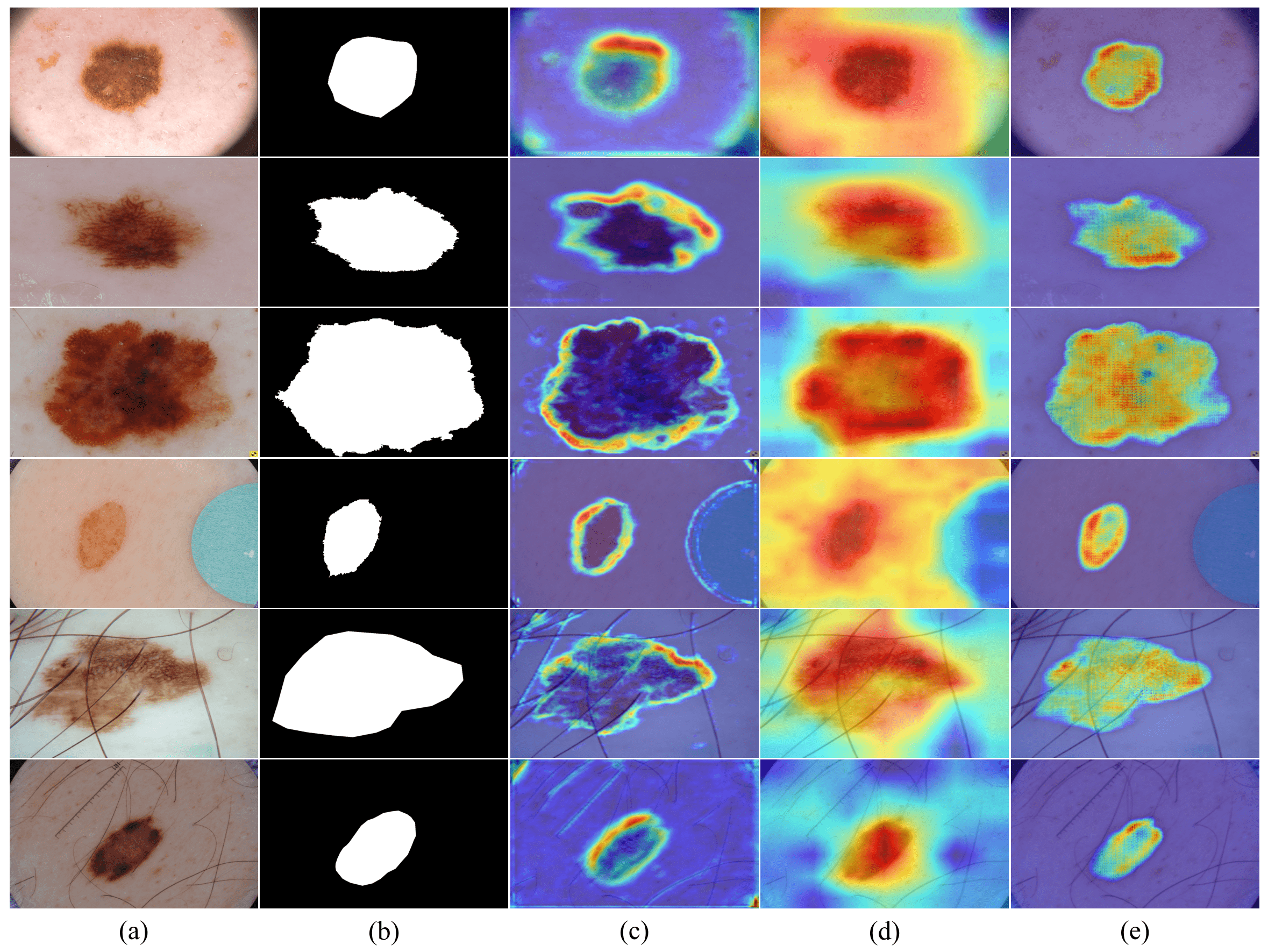}
  \caption{Visual comparison of different attention maps extracted by edge encoder and body encoder. (a) Input images. (b) Ground truth. (c) Attention map of the last stage in edge encoder. (d) Attention map of the last stage in body encoder.(5)Attention map of final layer of our model.}
  \label{encoder_visual}
\end{figure*}

In addition, we also perform a visual comparison of the segmentation results on the PH2 dataset among the UNet, AttUNet, FAT-Net, TransUNet, and LCAUnet. As shown in Fig. \ref{PH2_visual}, for most skin lesions with blurry boundaries and irregular shapes, our proposed LCAUnet still achieves the best segmentation results, demonstrating its effectiveness and robustness.

\subsection{Ablation studies}

To demonstrate the efficiency of different components in LCAUnet model, ablation experiments are conducted. We use a single-branch U-shaped architecture that includes only a body encoder and a decoder based on the simple concatenation of adjacent-scale features as our baseline. We then add PGMF module, EdgeEncoder, and LCAF module to the baseline network separately, resulting in the methods noted as baseline+PGMF, baseline+EdgeEncoder, and baseline+EdgeEncoder+LCAF. Comparative experiments are evaluated on these networks equipped with different modules on the ISIC2017 and ISIC2018 datasets.

Compared to the baseline, the baseline+PGMF and baseline+EE (EdgeEncoder) methods exhibite improvements of approximately 2.63$\%$ and 6.74$\%$, respectively in the IoU metric, and approximately 2.52$\%$ and 5.27$\%$, respectively in the Dice metric on the ISIC 2017 dataset, as shown in Table \ref{ablation_table}. These results demonstrate that PGMF facilitates the fusion of features between adjacent scales, while the edge information extracted by EdgeEncoder is crucial for improving segmentation performance. The baseline+EE+LCAF method further improve the IoU and Dice metrics by 0.63$\%$ and 1.02$\%$, respectively, on the ISIC 2017 dataset compared to the baseline+EE (EdgeEncoder) method. This indicates that the LCAF module can promote the fusion of edge and body information. By integrating all three modules, our method (baseline+EE+LCAF+PGMF) achieves the best performance, with approximately 2.01$\%$, 8.87$\%$, and 7.03$\%$ higher ACC, IoU, and Dice metrics than the baseline, respectively, on the ISIC 2017 dataset. Improvements in these metrics are also observed on the ISIC2018 dataset and PH2 dataset through similar experiments, and this further validates the effectiveness of the proposed modules.

To visualize the results of the ablation study, we select several models that integrate different modules and conducted a comparative visual analysis on the ISIC2017 dataset. The results depicted in Fig. \ref{ablation_visual} demonstrate a noticeable improvement to tackle complex segmentation challenges after incorporating the PGMF, EE(EdgeEncoder), and LCAF modules. Compared to methods that only integrate some of these modules on the baseline, LCAUnet model which integrates all three modules outperforms them and achieves the best segmentation results.

Furthermore, in order to investigate the ability of the edge encoder and body encoder to extract edge and body features separately, Grad-CAM \cite{selvaraju2017grad} is utilized to visualize the output feature maps of the final stage of both encoders. As shown in Fig. \ref{encoder_visual}, the attention region of edge encoder is mainly located at the edge of the skin lesion, which effectively extracted edge features. However, due to the locality of the convolutional operator, the edge encoder cannot accurately judge the semantics of each pixel based on global information, leading to extracted feature that may contain interference objects such as circular pieces and hair around the edge of the skin lesion. Meanwhile, body encoder effectively perceives the position of the body region by capturing long-range dependencies. While, since the transformer-based body encoder lacks the ability to capture local features, which makes poor  performance in edge segmentation. By integrating the advantages of the edge and body modalities, LCAUnet achieves superior segmentation of skin lesion edges while effectively perceiving the skin lesion region, as shown in Fig. \ref{encoder_visual} (e).

\section{Conclusions}

In this paper, we propose a novel skin lesion segmentation structure, namely LCAUnet, which extracts both edge and body features and integrates these complementary features to enhance the segmentation performance. Simultaneously, a local cross-modal fusion module LCAF that incorporates the edge and body modalities is constructed for feature fusion. Additionally, an PGMF module is employed in the decoder stage to better integrate features from adjacent scales. Comprehensive experiments are conducted on three public available datasets for model evaluation. Compared with existing methods, the LCAUnet model shows superior performance. It is capable of accurately segmenting irregular, boundary-blurred, and interference-present skin lesion regions, with good generalization performance. It is validated that the proposed LCAUnet model outperforms most of the SOTA methods. In future works, we will explore the universality and generalization of the model by applying it to other medical image segmentation challenges.


\bibliographystyle{elsarticle-num}
\bibliography{main}

\end{sloppypar}
\end{document}